\documentclass[aps,twocolumn]{revtex4}
\usepackage{amsmath}
\usepackage{graphicx}% Include figure files
\usepackage{dcolumn}% Align table columns on decimal point
\usepackage{bm}% bold math
\usepackage{epsfig}

\newcommand{\beq}{\begin{equation}}
\newcommand{\eeq}{\end{equation}}
\newcommand{\bee}{\begin{eqnarray}}
\newcommand{\eee}{\end{eqnarray}}
\newcommand{\ba}{\begin{eqnarray}}
\newcommand{\ea}{\end{eqnarray}}

\begin{document}
\title{Hydrodynamic interactions suppress deformation of suspension drops in Poiseuille flow}
\author{Krzysztof Sadlej}
\author{Eligiusz Wajnryb}
\author{Maria L. Ekiel-Je\.zewska}

\date{\today}

\address{%
Institute of Fundamental Technological Research,
Polish Academy of Sciences,
ul. Pawi{\'{n}}skiego 5B,
02-106 Warsaw,
Poland\\
}%

\begin{abstract}
Evolution of a suspension drop entrained by Poiseuille flow is studied numerically at a low Reynolds number. A suspension drop is modelled by a cloud of many non-touching particles, initially randomly distributed inside a spherical volume of a viscous fluid which is identical to the host fluid outside the drop. Evolution of particle positions and velocities is evaluated by the accurate multipole method corrected for lubrication, implemented in the {\sc hydromultipole} numerical code.  Deformation of the drop is shown to be smaller for a larger volume fraction. At high concentrations, hydrodynamic interactions between close particles significantly decrease elongation of the suspension drop along the flow in comparison to the corresponding elongation of the pure-fluid drop. Owing to hydrodynamic interactions,  
the particles inside a dense-suspension drop tend to stay for a long time together in the central part of the drop; later on, small clusters occasionally separate out from the drop, and are stabilized by quasi-periodic orbits of the constituent non-touching particles. 
Both effects significantly reduces the drop spreading along the flow. At large volume fractions, suspension drops destabilize by fragmentation, and at low volume fractions, by dispersing into single particles.
% left behind the drop. 
\end{abstract}

\maketitle

\section{Introduction}
Dynamics of a cloud of many non-touching micro-particles forming a suspension drop entrained by the Poiseuille flow
in a microchannel is interesting for various industrial, biological and medical applications, 
such as e.g.  transport in microfluidic devices~\cite{Sharp2005} or narrow channels~\cite{Davis2005}, drug or gene delivery through
the use of magnetic nano-particles~\cite{Dobson2006} or inhalator drug delivery~\cite{Edwards1997,Broday2003}.
The basic question, from a dynamical point of view, is how to control dispersion of the particles and the cloud deformation. 
This is important from the practical perspective, as in some aspects it is important that the particles or molecule clusters stay intact for a 
certain time or traveled distance, while in others fast dispersion is desired.

It is known from other contexts~\cite{NitscheBatchelor} that hydrodynamic interactions which arise between non-touching particles moving 
in a viscous fluid under a low Reynolds number, in general tend to keep close particles together in groups. 
This phenomenon is well studied on theoretical, numerical and experimental grounds~\cite{MMNS,EMG} for particle movement resulting from the gravitational field. In particular, an ensemble of initially randomly distributed particles on the average moves in the same way as a fluid drop settling down in a less dense host fluid. For a very long time, such a suspension drop behaves as a cohesive entity even though there is no surface tension to hold the suspended particles together.

The clustering effect has been also observed for groups made of several particles only. In gravitational field, three particles can stay together for a very long time, what gives an indication to the evolution characteristics of larger particle clouds~\cite{Janosi1997}. In analogy, periodic motion of two particles under shear flow~\cite{Bat07}
forms a benchmark which we expect to influence dynamics of large clouds of particles entrained by an ambient fluid flow. In both contexts, hydrodynamic interactions keep particles together for a long time. A similar mechanism has been observed in Ref.~\cite{Jones2001}, where stability of particle clusters of an ordered internal structure immersed in a shear flow was analyzed.

A cloud of randomly distributed particles in an ambient fluid flow, has yet not been investigated. 
This paper is therefore devoted to a study of the evolution of such a suspension drop entrained by a Poiseuille flow inside a two-wall channel, in the low-Reynolds-number regime. A suspension drop is modeled by a cloud of many non-touching particles, which are initially distributed randomly inside a given spherical volume of the fluid, identical to the host fluid outside. In the course of evolution, the particles are free to move relative to each other and the surrounding channel walls. 

The goal is to analyze to what extent hydrodynamic interactions between the particles inside a suspension drop influence its evolution, and in particular, its elongation and dispersion along the flow. The strength of the hydrodynamic interactions is tuned-up by the increasing volume fraction of the suspension. 

The structure of this paper is the following: section II contains the general system description followed by section III introducing briefly the numerical procedure used to calculate the hydrodynamic interactions. Section IV contains the main results and their discussion. The paper is summarized and concluded in section V.

\section{System}
Consider a suspension drop made of $N$ particles immersed in a viscous fluid identical to the fluid outside the drop. Each individual particle is a hard sphere of diameter $d$.  The stick boundary conditions on the particle surfaces are assumed.
The particles cannot overlap, and do not interact with each other through electrostatic or magnetic forces or other direct interactions. 
The particles are initially randomly distributed in a given spherical fluid volume of diameter $D$, with equal $N$-particle probability everywhere except overlapping. Volume fraction $\phi$ of the particles is therefore given as
\beq
 \phi = \frac{N d^3}{D^3}.
 \label{eq:phi}
\eeq

The fluid is bounded by two infinite planar walls separated by a distance $h$, which is much larger than the drop radius.
This configuration models a micro-channel of a width and length which are much larger than its height. Across the length of this channel a pressure gradient is exerted. In the absence of a drop, it would lead to the formation of a steady Poiseuille flow,
\beq
{\bf v}_0= 4 \text{v}_{m} \ z(h-z)/h^2\,\hat{\bf x}.
\label{eq:Poiseuille}
\eeq
\begin{figure}[h!]
\includegraphics[width=8.5cm]{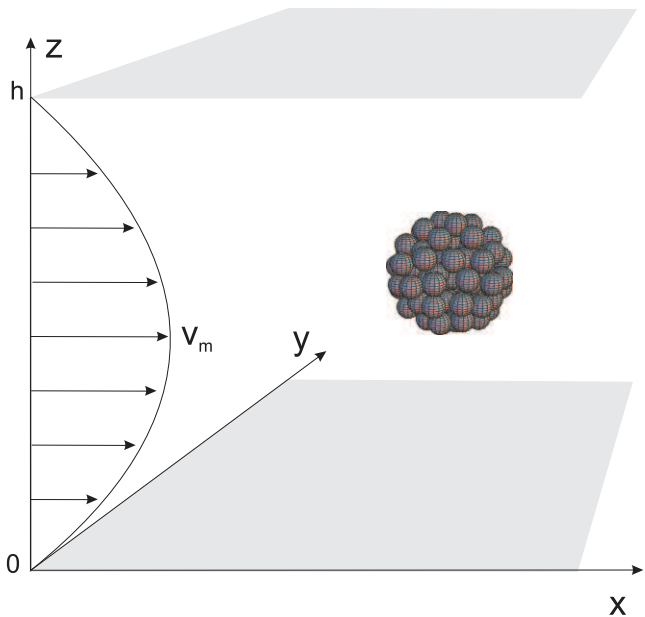}
\caption{Model system at initial moment of time.}\label{pict_model}
\end{figure}

As illustrated in Fig.~\ref{pict_model}, the suspension drop is immersed in the ambient flow~\eqref{eq:Poiseuille}, which modifies the fluid velocity and pressure fields, ${\bf v}({\bf r})$ and $p({\bf r})$. A low-Reynolds-number flow is assumed and  
described by the steady Stokes equations~\cite{Happel-Brenner:1986,Kim-Karrila:1991}, 
\beq
\eta\boldsymbol{\nabla}^2{\bf v} -\boldsymbol{\nabla} p  = \bm{0},\quad \boldsymbol{\nabla}\cdot{ \bf v}  = 0.
   \label{eq:stokes}
\eeq

Therefore, translational velocities of the particle centers, $d {\bf r}_i/d t$, $i=1,...,N$, are  linear functions of the maximal ambient flow velocity $\text{v}_{m}$, 
\beq
\frac{d {\bf r}_i}{d t} =\text{v}_{m}   \bm{C}_i(\bm{X}), 
\eeq
where the coefficients $\bm{C}_i(\bm{X})$ are Cartesian vectors depending on the configuration of all the particle centers, $\bm{X}=({\bf r}_1, {\bf r}_2, ... {\bf r}_N)$.

\section{Numerical procedure}
The Cartesian vectors $\bm{C}_i(\bm{X})$ are determined numerically by
the {\sc{hydromultipole}} algorithm, which implements the
theoretical multipole method of calculating
hydrodynamic interactions between bodies~\cite{Cichocki-Felderhof-Hinsen-Wajnryb-Blawzdziewicz:1994,Ekiel_Jezewska-Wajnryb:2008,Bhattacharya-Blawzdziewicz-Wajnryb:2005a,Bhattacharya-Blawzdziewicz-Wajnryb:2006a} within Stokesian dynamics. The defined cut-off parameter $L$
for the multipole method was chosen $L=2$ what means that $24$ multipole moments where calculated 
for each of the $N$ particles. Such a choice of the multipole cut-off is sufficient to achieve a 
precision of the calculated velocities, normalized by $\text{v}_m$, at least $5 \times 10^{-6}$ when calculating velocities of particles in our system. 
This precision estimate was found by calculating the velocity of the suspension drop at $t=0$ for $L$ ranging from
2 to 8 with $\phi=40\%$ (dense drops, i.e. particles packed close together).

The particle-wall interactions are incorporated by the single-wall superposition
of hydrodynamic forces for the two-wall system~\cite{Bhattacharya-Blawzdziewicz-Wajnryb:2005a}. This approximation is poor for narrow channels when
the initial diameter of the drop is comparable to the width of the channel but
it performs well for wide channels~\cite{Bhattacharya-Blawzdziewicz-Wajnryb:2005a}. Here we consider the case of relatively wide channels
compared to the drop diameter when use of the single-wall superposition approximation
is fully justified.

The first analysis performed was the estimation of wall effects encountered. We compared the initial mean velocities of identical drops (same initial configuration of particles) in the same external Poiseuille flow, using numerical {\sc hydromultipole} procedures for an unbounded fluid and fluid between two parallel walls. 
At the initial moment of time, the difference between the two calculated drop velocities, normalized by $\text{v}_m$, was of the order of $2 \times 10^{-6}$, i.e. smaller that the computational accuracy. 
Taking into account the above result, we  used the numerical {\sc hydromultipole} codes for an unbounded fluid, and benefited from the three-time increase in calculation speed.

Having calculated the instantaneous velocities of all particles, the evolution of the drop was
determined by time stepping the set of coupled differential equations for
each particle position.

The number of independent simulation runs finally performed varied with the volume fraction. For $\phi$ up to $30\%$, ten independent initial configurations where considered. For volume fractions $40\%$ and $50\%$ twenty simulation runs where performed. This was motivated by larger  
fluctuations observed for larger volume fractions.

\section{Evolution of a suspension drop}
\subsection{Model parameters}
The simulated model is described by three parameters, the channel height $h$, the number of particles in the suspension drop $N$ and the volume fraction $\phi$. 
The number of particles $N$ in the drop was held constant, equal to $N=80$. Our goal was to study how the drop evolution would change with the increased volume fraction, when the hydrodynamic interactions between the close particles are enhanced. Increasing volume fractions, we also increased the size of the particles, $d$, according to Eq.~\eqref{eq:phi} with a constant drop diameter $D$.
At the same time, we kept a constant channel height $h$, so as the ratio $D/h$ was also held constant.  This ensured that the flow gradient over a drop diameter was approximately equal for all volume fractions. Suspension drops of different volume fraction (i.e different diameter when composed of equal number of particles) could then be compared in terms of the influence the same flow had on the structure of the drop. 
We chose a low value $D/h=0.117$, with the corresponding values of $h$ listed in Table~\ref{Tab1}.
This enabled us to make hydrodynamic interactions of the drop with the walls very weak, and therefore to focus on the hydrodynamic interactions between particles inside the drop.

\begin{table}[h!]
\caption{The channel height $h/d$ used in our simulations for a given volume fraction $\phi$, with $N=80$ and $D/h=0.117$.}
\begin{tabular}{|c|c|}
\hline
$\phi$ & $h/d$\\
\hline\hline
5\%  & 100.00 \\
10\% & 79.37 \\
20\% & 63.00\\
30\% & 55.04 \\
40\% & 50.00  \\
50\% & 46.42 \\
\hline
\end{tabular}
\label{Tab1}
\end{table}

From now on, we normalize distances by the height $h$ of the channel (i.e. the distance between the channel walls).  Velocities are normalized by the maximum velocity of the flow, $\text{v}_m$, attained in the middle of the channel. Time $t$ in the simulations is measured in  $h/\text{v}_m$. 
All simulations were performed for times approximately up to $t=300$. 

\subsection{Initial moment of time}
\begin{figure}[h!]
\includegraphics[width=8.5cm]{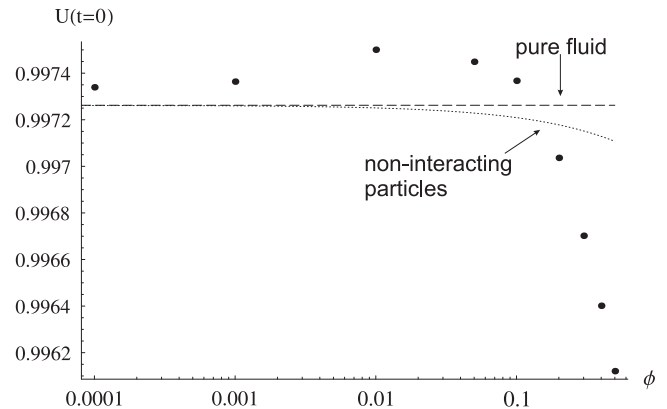}
\caption{The dimensionless velocity of a suspension drop at $t=0$ averaged over $100$ independent, random initial conditions (dots). Errors are smaller than the size of the dots. The dashed line is the limiting case of a  volume of fluid having the same size and initial position as the considered suspension drops. The dotted line corresponds to a suspension drop composed of non-interacting particles.}
\label{U_avr}
\end{figure}

The average velocity of a suspension drop 
\beq
 U =\frac{1}{M}\sum_{j=1}^{M}\frac{1}{N}\sum_{i=1}^{N} U_{i(j)},
 \label{eq:U_avr_t0}
\eeq
calculated at $t=0$, is plotted in Fig. \ref{U_avr} as a function of volume fraction. Here $N=80$ is the number of particles in a drop and $M=100$ is the number of independent simulations performed for different random configurations of the particles, and $U_{i(j)}$ is the velocity of particle $i$ in simulation $j$, divided by $\text{v}_m$.  Errors of the average velocity(Eq. \ref{eq:U_avr_t0}) are smaller than the size of the plotted points. 

In comparison, a spherical volume of pure fluid identical to the analyzed suspension drop moves at the initial moment of time with a velocity
\beq
 U_f = \frac{1}{\Omega}\int_{\Omega} \tilde{\text{v}}_0(z) d\Omega = \left(1-\frac{D^2}{5 h^2}\right),
 \label{eq:U_analytic}  
\eeq
where $\Omega$ denotes the volume of a sphere of diameter $D$ centered at $(0,0,1/2)$ and $\tilde{\text{v}}_0(z)\!=\!|{\bf v}_0|/\text{v}_m\!=\!4z(1\!-\!z)$. This mean equals $U_f= 0.997262 $ for $D/h=0.117$ and is denoted by the dashed line in Fig.~\ref{U_avr}.

The dotted curve is the velocity of a suspension drop composed of non-interacting particles, i.e. of
a volume of fluid, corrected by the Faxen term, corresponding to the non-zero diameter $d$ of particles,
\beq
 \delta U = \frac{d^2}{24h^2}\nabla^2 \tilde{\text{v}}_0 =  - \frac{1}{3}\frac{d^2}{h^2}.
 \label{faxen}
\eeq
Therefore, according to Table~\ref{Tab1}, the Faxen correction depends on volume fraction. For example, $\delta U=1.3 \times 10^{-4}$ for $\phi=0.40$ (i.e. $h=50d$).

It can be noticed in Fig.~\ref{U_avr} that a change of the suspension volume fraction results in an overall small difference in mean velocity of the suspension drop, at most of the order of $10^{-3}$. Nevertheless, the changes in evolution with the increase of volume fraction are substantial, as will be pointed out in the next section.

\subsection{Snap-shots from suspension drop evolution}
Although the difference in the mean velocity of the suspension drop changes only slightly with the 
\begin{figure*}[!]
\includegraphics[width=6cm]{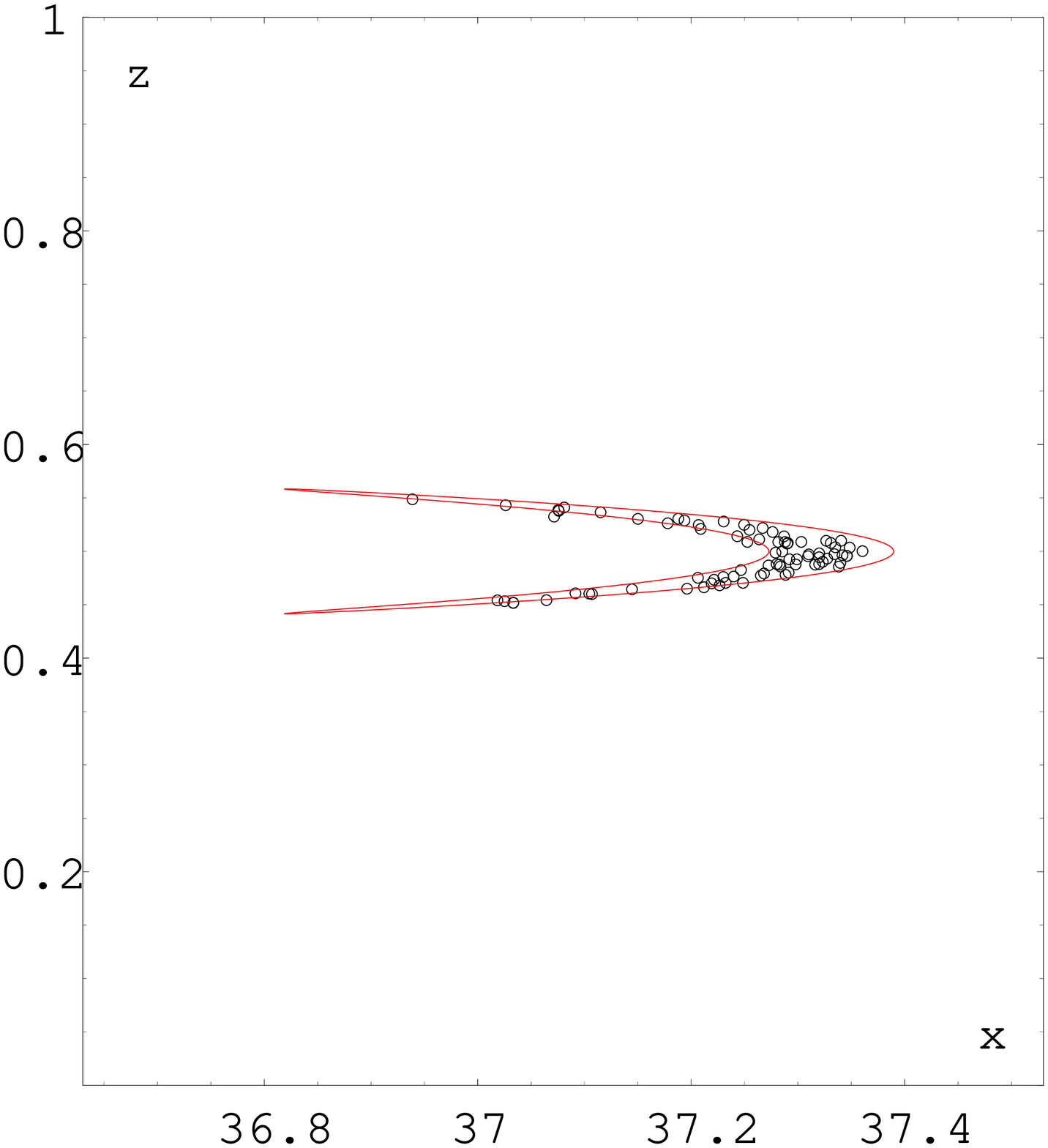}
\includegraphics[width=6cm]{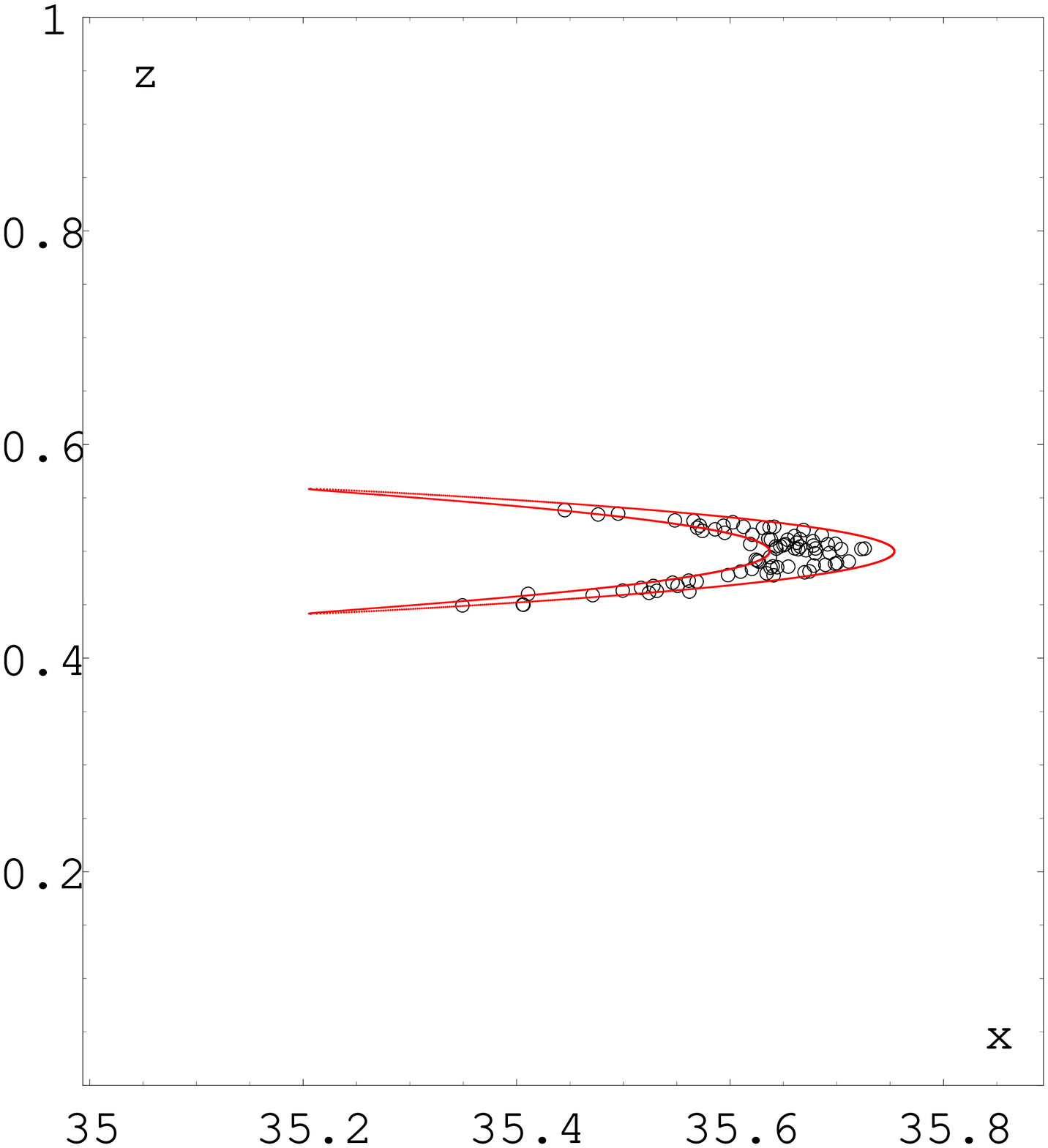}
\includegraphics[width=6cm]{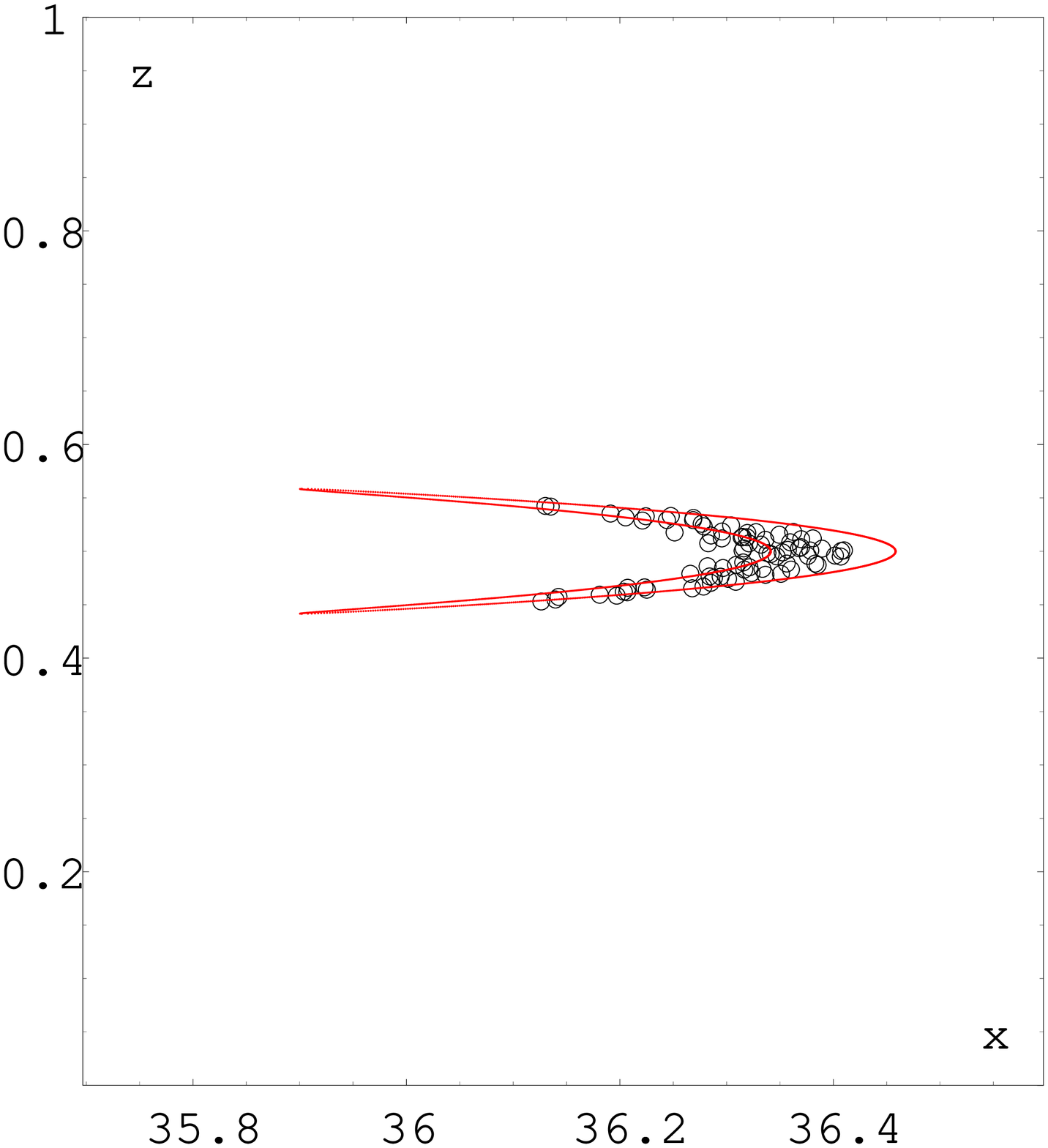}
\includegraphics[width=6cm]{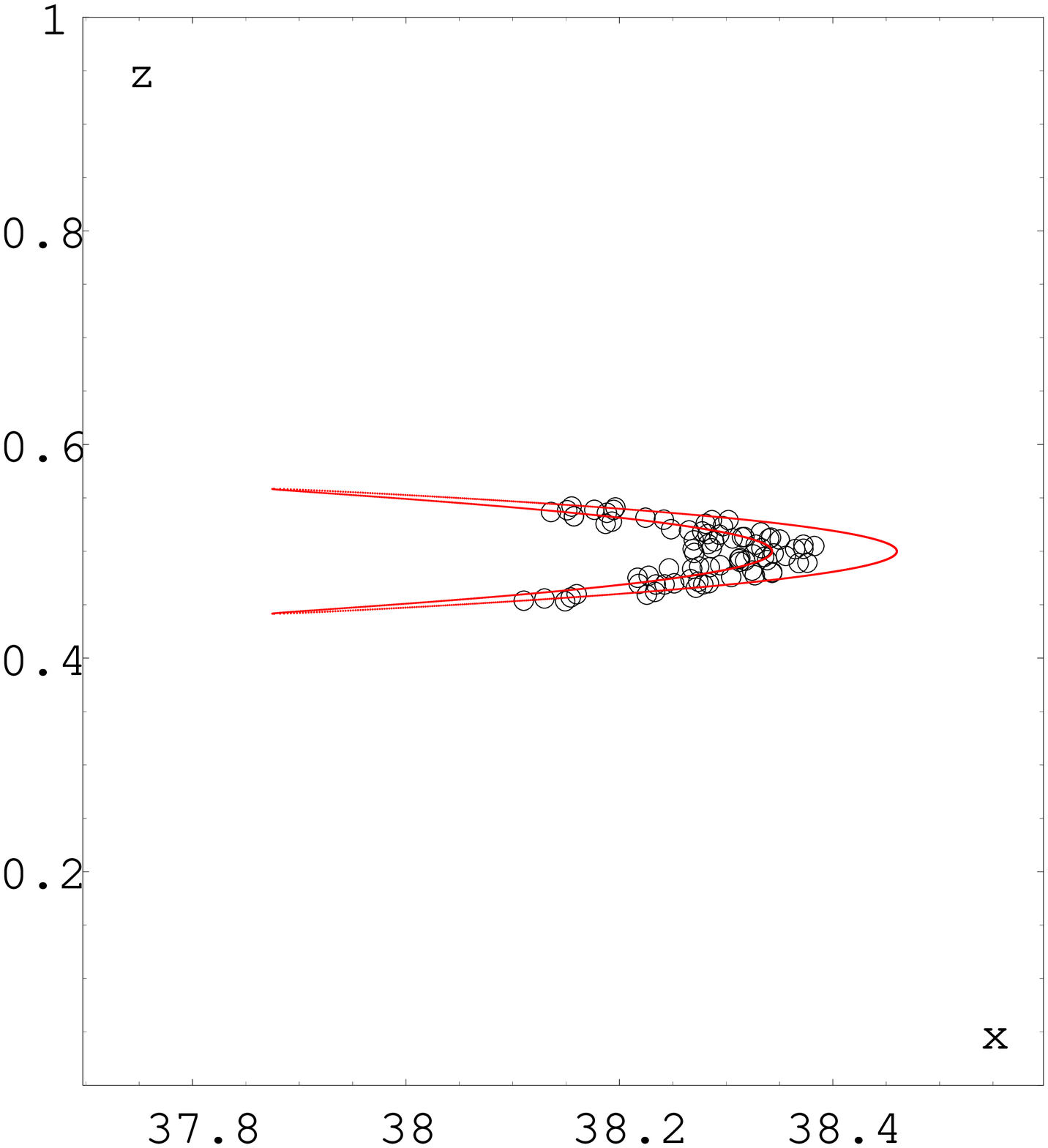}
\includegraphics[width=6cm]{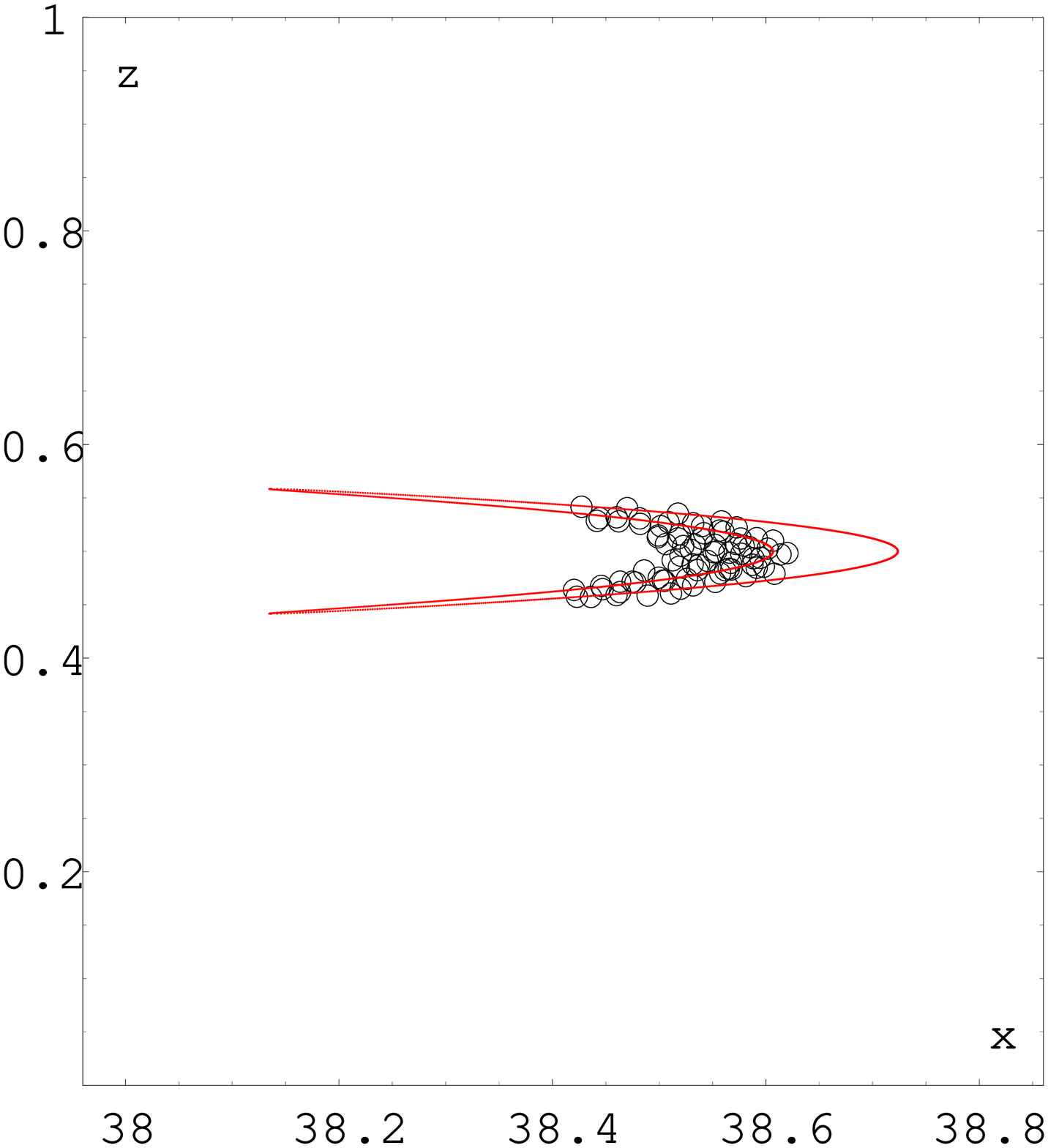}
\includegraphics[width=6cm]{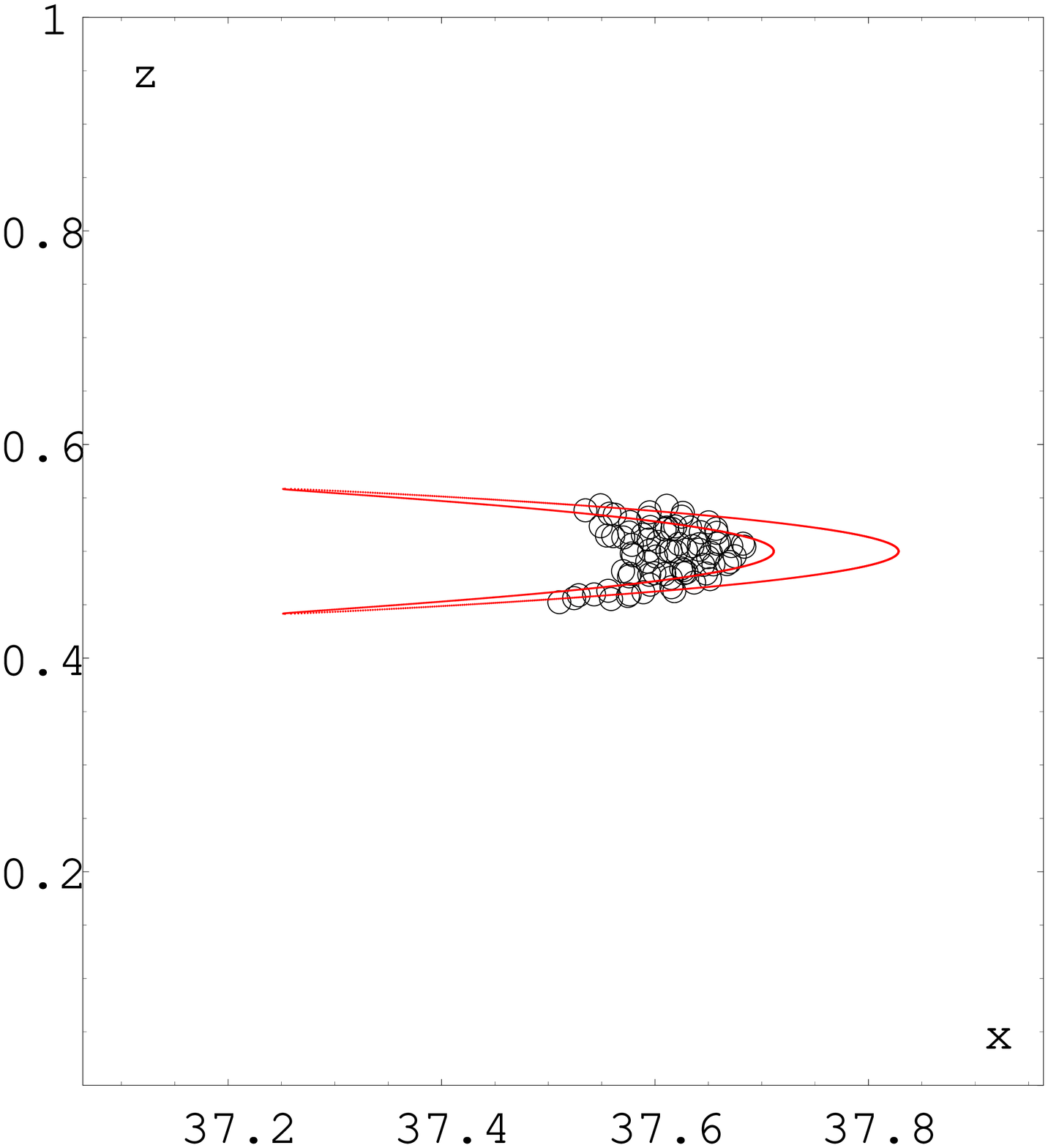}
\caption{Snap-shots from drop evolution simulations \cite{movie03a,movie03b,movie03c,movie03d,movie03e,movie03f}. All particles have been projected onto the $y=0$ plane. 
The red outline (given by Eq. (\ref{curve_xz})) is the circumference of the projection of the instantaneous position of the fluid volume, which would be initially identical with the suspension drop.(enhanced online)}\label{fig_evo_fi}
\end{figure*}
\begin{figure*}[!]
\includegraphics[width=6cm]{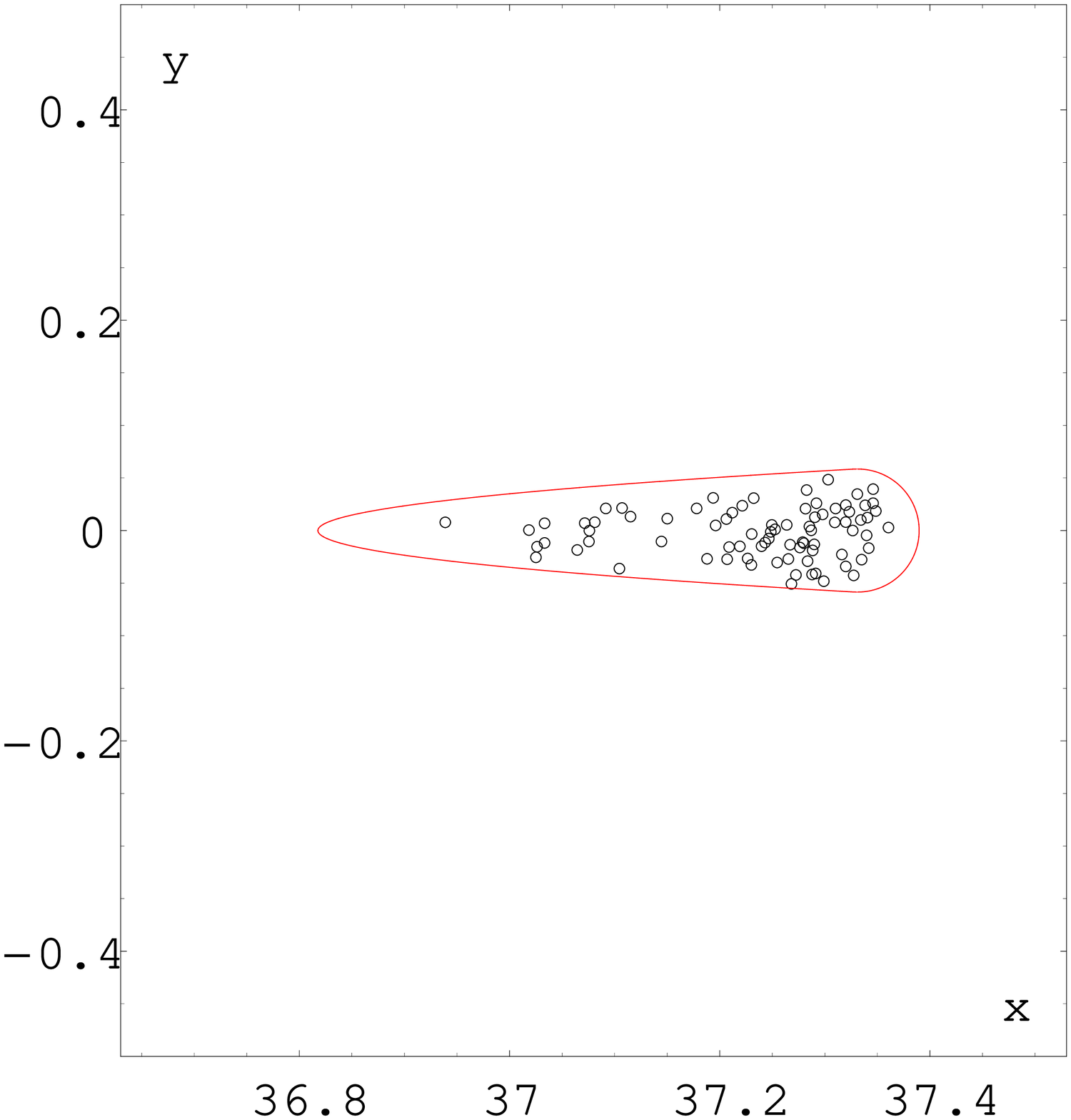}
\includegraphics[width=6cm]{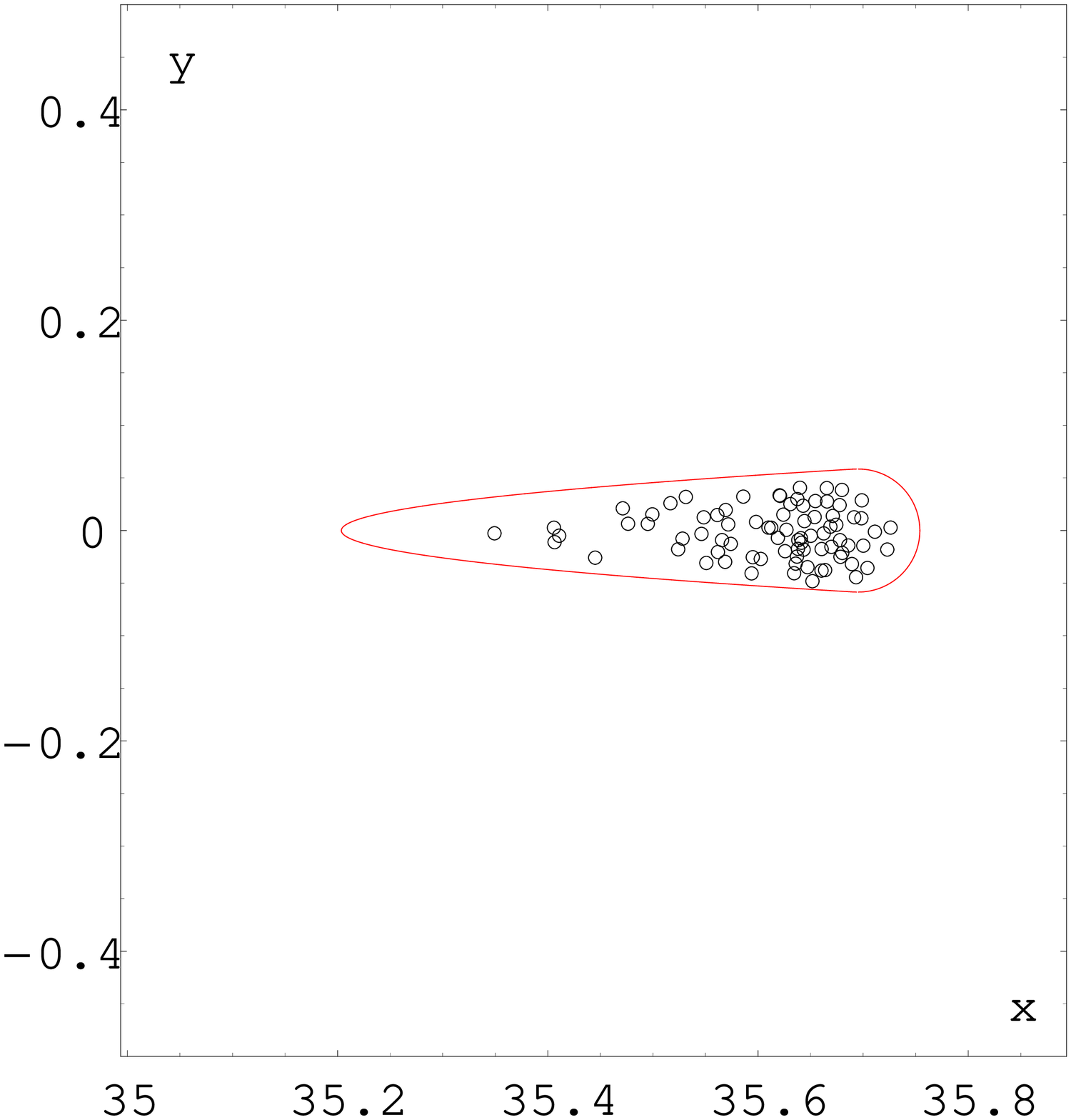}
\includegraphics[width=6cm]{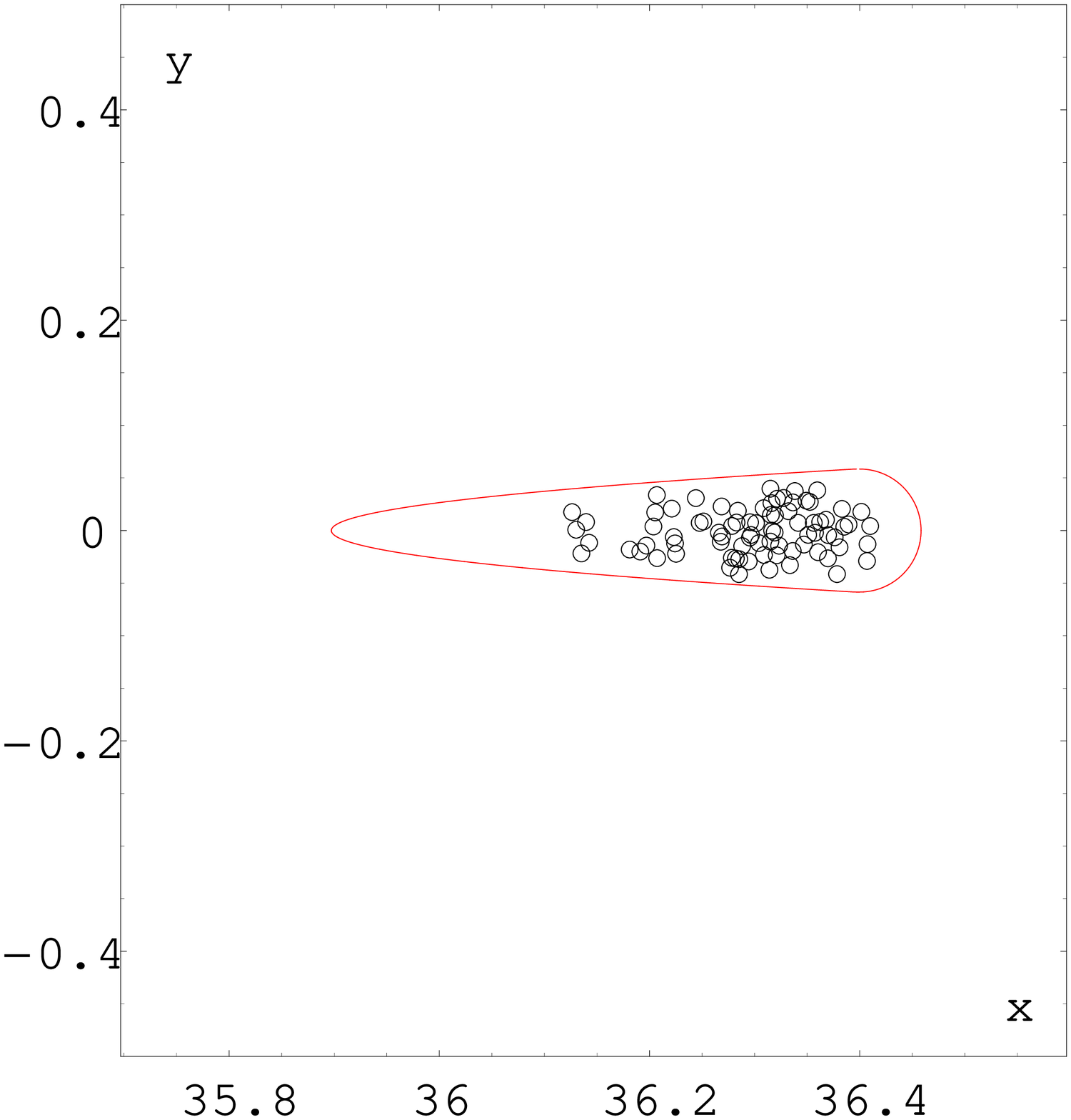}
\includegraphics[width=6cm]{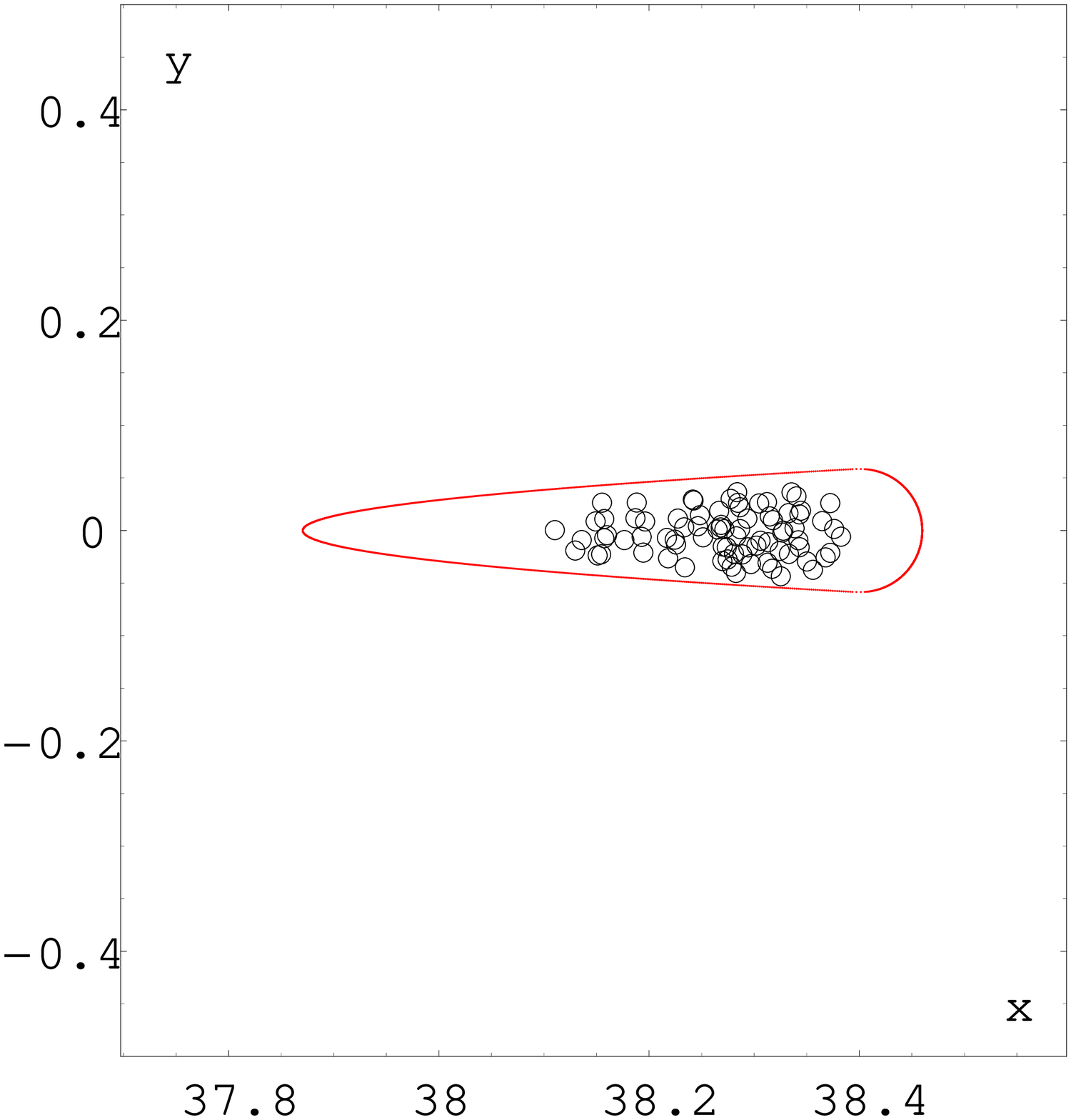}
\includegraphics[width=6cm]{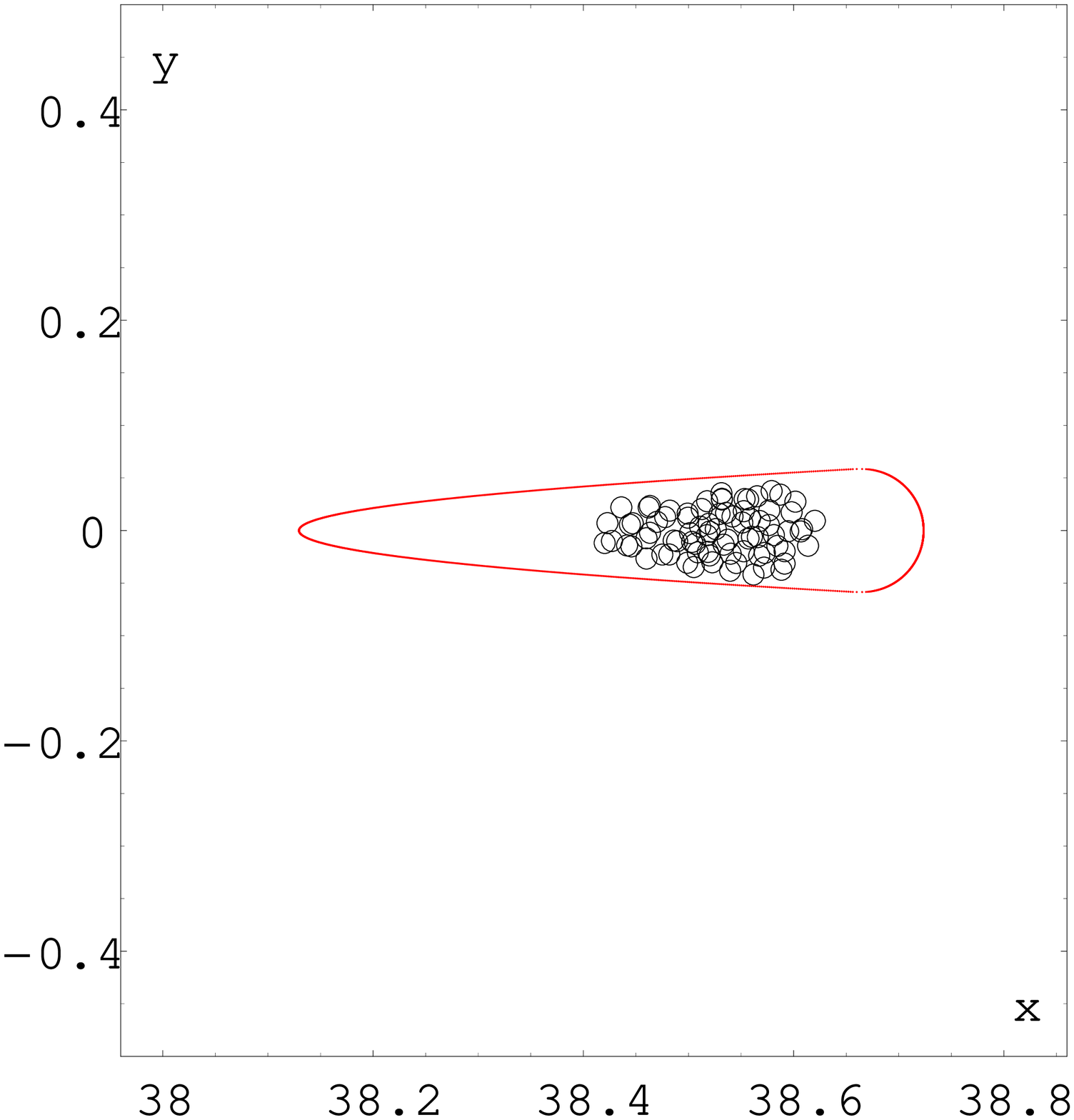}
\includegraphics[width=6cm]{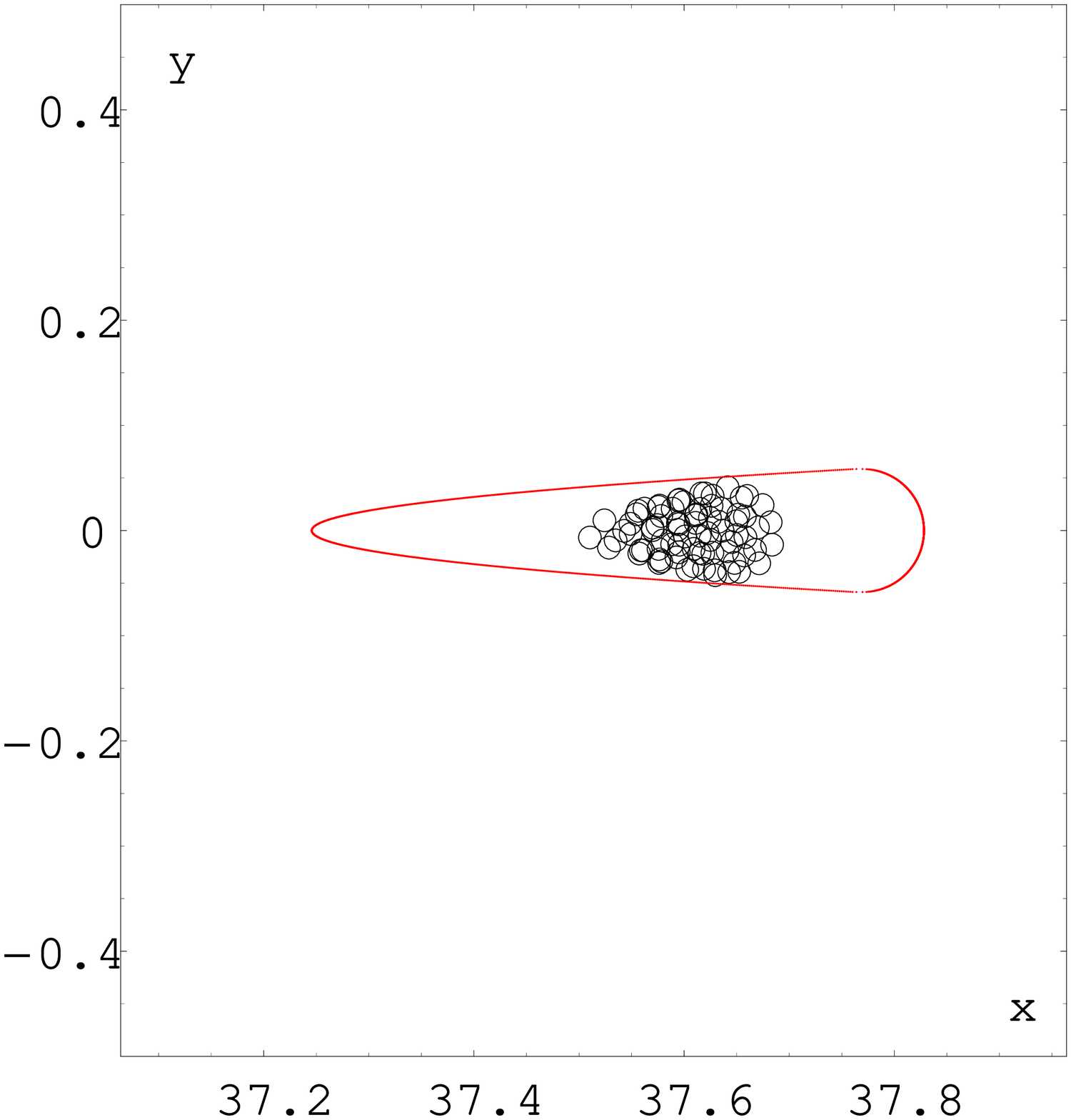}
\caption{Snap-shots from drop evolution simulations \cite{movie04a,movie04b,movie04c,movie04d,movie04e,movie04f}. All particles have been projected onto the $z=1/2$ plane. 
The red outline (given by Eqs. (\ref{curve_xy_1}-\ref{curve_xy_22})) is the circumference of the projection of the instantaneous position of the fluid volume, which would be initially identical with the suspension drop. (enhanced online)}\label{fig_evo_fi_xy}
\end{figure*}
change of its volume fraction, a quick look at the simulation snapshots (Fig. \ref{fig_evo_fi}, see \cite{movie03a,movie03b,movie03c,movie03d,movie03e,movie03f} and Fig. \ref{fig_evo_fi_xy}, see \cite{movie04a,movie04b,movie04c,movie04d,movie04e,movie04f}) shows a clear change in the behavior of all particles. As the volume fraction increases, particles tend to stay clustered together longer. The stretching of the initial shape of the suspension drop therefore highly depends of its volume fraction. In order to show this phenomenon  clearly it is instructive to compare the positions of particles with the evolved shape of an initially spherical volume of fluid. The equation describing the circumferential surface of this fluid volume is
\bee
   (x-\tilde{\text{v}}_0(z) t)^2 &=& \frac{D^2}{4h^2}-z^2 - y^2,\label{sd}\\
\tilde{\text{v}}_0(z)&=&4z(1-z).\nonumber
\eee

As simulations shown and discussed in this paper are presented in projections ($y=0$ or $z=1/2$), the interior of the drop contained inside the above surface has to be accordingly projected. Therefore,
\begin{enumerate}
\item The boundary of the fluid-drop projection on the $y=0$ plane is given by Eq.~\eqref{sd} with $y=0$.
\item The fluid-drop projection on the $z=1/2$ plane is a superposition of circles shifted with respect to each other. Its boundary is therefore more sophisticated, see  Eqs.~\eqref{curve_xy_1}-\eqref{curve_xy_22} derived in Appendix~\ref{AA}.
\end{enumerate}

The boundaries of the fluid-drop projections on the $y=0$ and $z=1/2$ planes are plotted in  
Figs. \ref{fig_evo_fi} and \ref{fig_evo_fi_xy}, respectively, and compared with the snapshots of the underlying movies [click on figure to watch movie], presenting the suspension drops at the same time instant. The scale on the $x$, $y$ and $z$ axes is the same, to keep spherical shape of the particles and the initial volume of the drops.

For small volume fractions 
the particles move in a similar way as the pure fluid. 
Particles initially concentrated in a spherical drop get spread out evenly and the initial shape of the drop is clearly stretched resembling the evolved shape of the corresponding spherical volume of pure fluid. No group formation is observed and hydrodynamic interaction between particles seems to be minor. 

As the volume fraction increases, the shape of the suspension drop at a given time instant 
is clearly less and less elongated by the flow than the evolved reference volume of pure fluid, as shown in Figs.~\ref{fig_evo_fi}-\ref{fig_evo_fi_xy}. Notice that for the largest volume fraction $\phi=50\%$, the drop's shape is still quite close to a sphere.

As the volume fraction increases, the evolution of the suspension drop changes significantly, and the  particles have a tendency to stay clustered in a single group for a longer time, as illustrated in the movies linked to Figs.~\ref{fig_evo_fi}-\ref{fig_evo_fi_xy}. 
 For larger times,  deformation of a dense suspension drop increases, and small groups of particles separate out from the main cluster. 
Particle groups which form in both tails, rotate relative to their center of mass, due to the gradient of the flow, in a similar way as two particles in a shear flow~\cite{Bat07}. The average number of the particles in such a group increases with time and is larger for a higher volume fraction.

\subsection{Quantitative analysis}
By viewing the simulation results one can conclude that hydrodynamic interactions hold particles together more effectively, when these are packed in larger volume fractions. This qualitative remark can be quantified by analyzing 
the time-dependent dispersion of the particle positions inside a drop, averaged over $M$ simulations corresponding to different random initial configurations. 
The dispersion along $x$-direction is defined by the following formula,
 \beq
   \sigma_x = \frac{1}{M}\sum_{j=1}^{M} \sqrt{\frac{1}{N}\sum_{i=1}^{N} x_{i(j)}^2 - \left(\frac{1}{N}\sum_{i=1}^{N}x_{i(j)}\right)^2},
   \label{eq:sigma_avr}
\eeq
where $N=80$ is the number of particles in a drop and $M$ is the number of simulations performed. The $x$-coordinates $x_{i(j)}$ of each of the $i=1,\ldots, N$ particles in each of the $j=1, \ldots, M$ simulations are functions of time, and so is the dispersion itself. The same formula leads to the definition of $\sigma_y$ and $\sigma_z$ once $x_{i(j)}$ is exchanged for the appropriate coordinate.

\begin{figure}[h!]
\includegraphics[width=8.5cm]{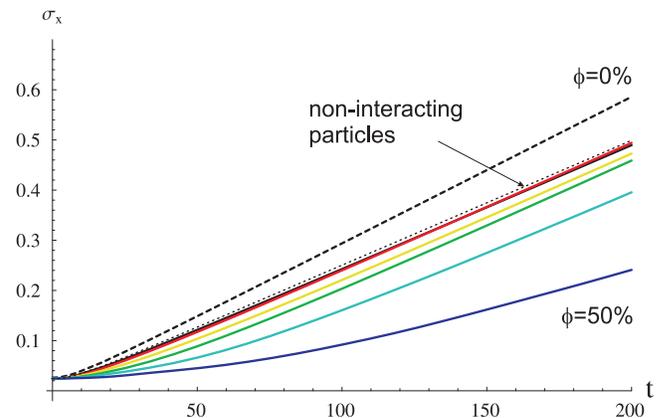}
\caption{Dispersion of particle positions in the drop along the flow (x) direction, evolving in time. Results averaged over all tested initial conditions. Color-coding see Table \ref{tab_color}; the larger volume fraction, the smaller is $\sigma_{x}$. The dashed line ($\phi\!=\!0\%$) corresponds to  
pure fluid, Eq.~\eqref{sigmaf}, and the dotted one to the non-interacting particles, Eq.~\eqref{ni}.} 
\label{fig_S_avr}
\end{figure}

Fig. \ref{fig_S_avr} shows the dispersion $\sigma_{x}$ of particle positions in the drop along the flow (x) direction, evolving in time, for different volume fractions. 
Color-coding, representing different volume fractions, has been chosen according to Table \ref{tab_color}. At a given time instant,  the dispersion $\sigma_{x}$ is smaller for a larger volume fraction $\phi$, and this effect is significant.  
 
\begin{table}[h!]
\caption{Color-coding used to differentiate results for various volume fractions}
\label{tab_color}
\begin{tabular}{|c|c|}
\hline
color & $\phi$ \\
\hline\hline
 black & $5\%$\\
 red & $10\%$\\
 yellow & $20\%$\\
 green & $30\%$\\
 light blue & $40\%$\\
 dark blue & $50\%$\\
\hline
\end{tabular}
\end{table}

For comparison, the dashed line in Fig.~\ref{fig_S_avr} is the analytical result calculated for a spherical volume of pure-fluid evolving with the flow ($\phi=0\%$). Its dispersion was calculated as
\begin{eqnarray}
\sigma_{f,x}&=& \sqrt{\frac{1}{\Omega}\int_{\Omega}\left(x+\tilde{\text{v}}_0(z)t-\bar{x}\right)^2 d\Omega}
\nonumber\\
&=&
\frac{1}{10} \sqrt{\frac{32 t^2 D^4}{7 h^4}+\frac{5 D^2}{h^2}},\label{sigmaf}
\end{eqnarray}
where
\beq
 \bar{x} = t  \left(1-\frac{D^2}{5 h^2}\right).
\eeq
The dispersion $\sigma_x$ of a pure-fluid drop is of course larger than the dispersion of a suspension drop. To investigate the effect of the excluded volume, the dispersion $\sigma_{x,ni}$ of a suspension drop made of fictitious non-interacting particles is also evaluated, defined by Eq.~\eqref{eq:sigma_avr} with the real particle positions $x_{i(j)}(t)$ replaced by $x_{i(j),ni}(t)$, 
the time-dependent positions of the fictitious particles, which would not interact hydrodynamically, and just translate along the Poiseuille fluid flow, $\tilde{v}_0$, with the Faxen velocity, $\tilde{v}_0+\delta U$, and the Faxen correction $\delta U$ given by Eq.~\eqref{faxen}, 
\bee
x_{i(j),ni}(t) &=& x_{i(j),ni}(0) + (\tilde{v}_0+\delta U) t.\label{ni}
\eee
The initial configurations of the fictitious particles are the same as those used in the suspension-drop evolution, 
$x_{i(j),ni}(0) = x_{i(j)}(0)$.
The result is plotted in Fig.~\ref{fig_S_avr} with a dotted line. The curves corresponding to $\phi=5\%,\;40\%$ and $50\%$ are practically superimposed. Comparison with the pure fluid and 
with the suspension drop indicates that elongation of the suspension drop is significantly suppressed by strong hydrodynamic interactions between particles at larger volume fractions.

\begin{figure}[h!]
\includegraphics[width=8.5cm]{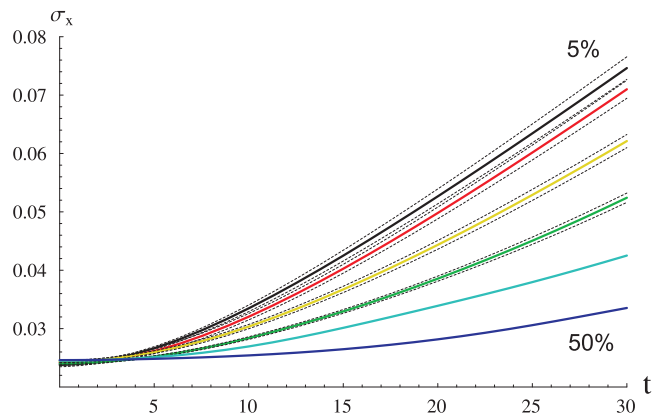}
\caption{Dispersion of particle positions in the drop along the flow (x) direction evolving in time. Results averaged over all tested initial conditions. Top to botton: $5\%$ to $50\%$ respectively. Color-coding see Table \ref{tab_color}.  The dotted lines show the standard mean errors. For $40\%$ and $50\%$ the errors are smaller than the width of the curve.}\label{fig_S_t30}
\end{figure}

In Fig. \ref{fig_S_t30} a closeup of Fig. \ref{fig_S_avr} for short times is shown. Further, the mean standard errors have been plotted (dotted lines) together with the original curves in order to show that the results obtained distinguish between different volume fractions. For volume fractions $40\%$ and $50\%$ the errors are smaller than the width of the curve. 

\begin{figure}[h!]
\includegraphics[width=8.5cm]{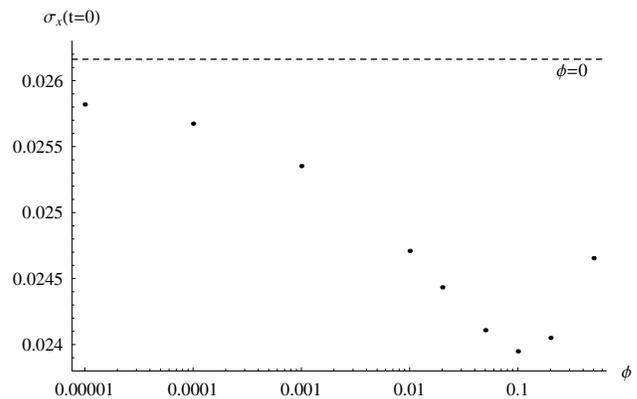}
\caption{Dispersion of particle positions along the flow (x) direction calculated at $t=0$.  The dashed line is the limiting analytical solution. Error-bars are smaller then the size of the points.}\label{fig_S_t0}
\end{figure}

Systematically, for a larger volume fraction, the dispersion $\sigma_{x}$ is smaller, if time is not too small, e.g. $t>5$. 
Notice that at $t=0$ the curves plotted in Fig. \ref{fig_S_t30} do not overlay each other. On first notice this might seem wrong. But when given further insight, the dispersion at $t=0$ is shown to change with the volume fraction, as plotted in Fig. \ref{fig_S_t0}. 
This graph shows the dispersion of particle positions along the flow (x) direction calculated at $t=0$ for volume fraction ranging from $0.001\%$ to $50\%$. The expected error of $\sigma_x$ is equal the standard deviation of the dispersion within a single drop, divided by $\sqrt{M}$. For $t=0$, the number of initial configurations randomly selected is $100000$ and thus the error-bars are smaller then the size of the points. The non-monotonic dependence of the drop dispersion on the volume fraction is a strictly statistical effect due to excluded volume. In particular, notice that the dispersion for $5\%$ is smaller than for $50\%$ as also visible on Fig. \ref{fig_S_t30}. The dashed line is the limiting analytical solution at $t=0$ for a volume of pure fluid (i.e. $\phi=0$). Of course $\sigma_y(t=0)=\sigma_z(t=0)=\sigma_x(t=0)$.
We checked that for a larger number of particles in the suspension drop (and therefore for a smaller particle diameter $d$), the dispersion $\sigma_x$ slightly increases.

For larger times, the value of the drop dispersion in the flow direction decreases more than by a factor of 2, for volume fractions changing form $5\%$ to $50\%$. This gives a clear indication that hydrodynamic interactions tend to hold close particles together, and therefore suppress deformation of the drop along the flow. 

\begin{figure}[h!]
\includegraphics[width=8.5cm]{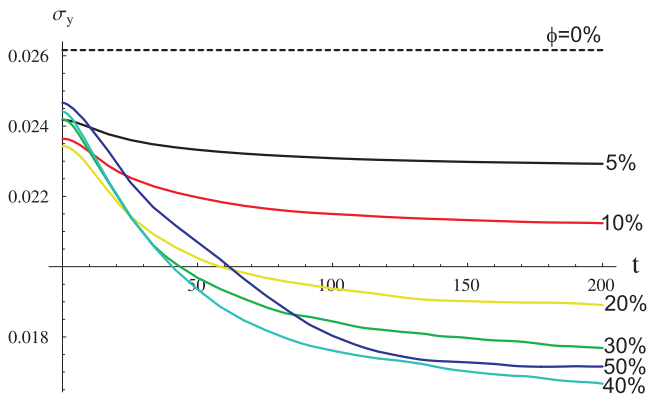}
\caption{Dispersion of particle positions in the $y$ direction, evolving in time. Results have been averaged over all tested initial conditions. The dashed line is the limiting result for a pure-fluid drop. The dotted lines show the standard mean errors.}\label{fig_S_y}
\end{figure}

\begin{figure}[h!]
\includegraphics[width=8.5cm]{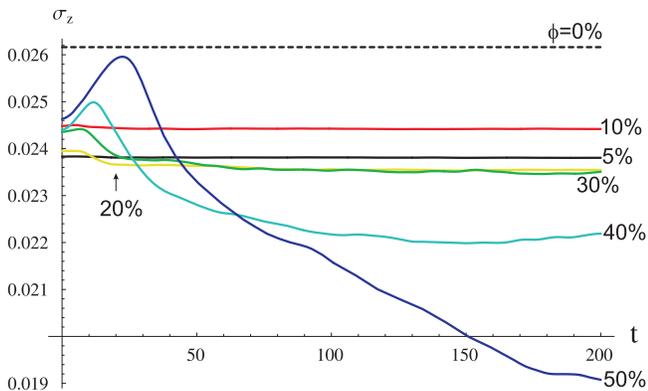}
\caption{Dispersion of particle positions in the $z$ direction, evolving in time. Results averaged over all tested initial conditions. The dashed line is the limiting result for a pure-fluid drop. The dotted lines show the standard mean errors.}\label{fig_S_z}
\end{figure}

Dispersion of particle positions along the transverse (y) and (z) direction, respectively, evolving in time, is shown in the two last figures in this section - Fig. \ref{fig_S_y} and Fig. \ref{fig_S_z}. Results have been averaged over all tested initial conditions. 

For comparison, we show also the corresponding dispersion of the pure fluid,
\bee
\sigma_{f,y} &=& \sigma_{f,z} = \sigma_{f,x}(t=0)=0.02616.
\eee
In the transverse directions, the dispersion of the pure-fluid drop  is constant in time.

 Note that in the direction  transverse to the flow the drop gets contracted as it evolves. This effect is small, but clearly visible as the volume fraction of the drop increases. It is also caused by hydrodynamic interactions between close particles.

\subsection{Group formation}
\label{sec:group}
When analyzing the simulation results at longer times, it becomes clear that for larger volume fractions, 
the particles which are lost from the main drop have a tendency to form small groups in the course of further evolution. This effect is readily visible in the movies linked to Figs.~\ref{fig_evo_fi}-\ref{fig_evo_fi_xy}, and it is a clear indication of hydrodynamic interactions between close particles. 

In particular, for simulations with $\phi\geq 30\%$ groups of several particles where observed to stay together until the end of the simulation at $t=300$. During this time each group rotated and particles interchanged places. This behavior of close particles resembles periodic trajectories of two particles in shear flow \cite{Bat07}.

We decided to check the group formation quantitatively by introducing an algorithm working in the following way: First, all particles are divided into two ensembles, depending if their position $z>1/2$ or $z \le 1/2$. Then, within these two ensembles, positions of their centers in the flow direction are compared by ordering. If between any two consecutive particle centers in such a sequence, the distance in the (x) flow direction is larger than a dimensional parameter $g$, then these particles are said to belong to two distinct groups. This grouping parameter if fully arbitrary. We chose $g=3d$. Different choices where checked - this discussion can be found in  Appendix~\ref{AB}.

\begin{figure}[h!]
\includegraphics[width=8.5cm]{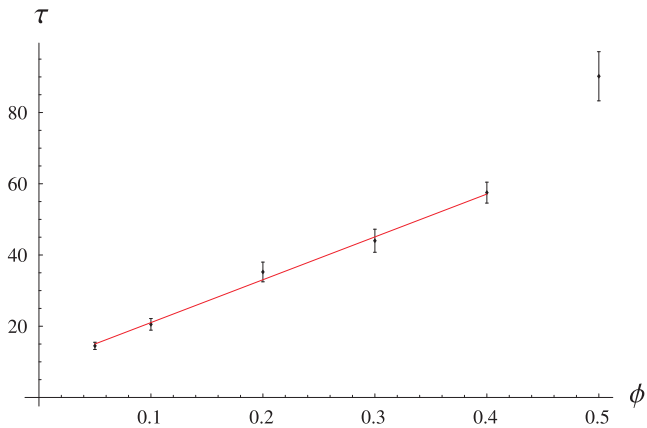}
\caption{Averaged time $\tau$ of first destabilization as a function of the drop volume fraction $\phi$. Error-bars are equal the 
standard deviation of the average value. The line is a linear fit to the data $5\%\leq\phi\leq40\%$. }\label{fig_destab}
\end{figure}

Having in mind the definition of a group, the averaged time $\tau$ of a first destabilization defined as formation of at least two groups of particles is evaluated and shown in Table \ref{tab01} and Fig. \ref{fig_destab}. Clearly, for a larger volume fraction, hydrodynamic interactions  between closer particles keep them together in a single cluster for a longer time.

\begin{table}[h!]
\caption{Averaged time $\tau$ of first destabilization defined as formation of at least two groups of particles.}
\label{tab01}
\begin{tabular}{|c|c|}
\hline
$\phi$ & $\tau$ \\
\hline\hline
5\%  & 14.5\\
10\% & 20.5\\
20\% & 35.2\\
30\% & 44.0\\
40\% & 57.0\\
50\% & 90.0\\
\hline
\end{tabular}
\end{table}

The error-bars in the figure correspond to the standard deviation of results calculated for the individual simulations. The line is a fit to the data $5\%\leq\phi\leq40\%$ given by the formula
  $\tau = 9.0 + 120.3 \phi$. The point for $\phi=50\%$ is substantially above this fit, while all other results seem to be well in accordance with the linear behavior.

The average number of distinct groups which are formed after time $t=50,100,150,200$ is listed in Table~\ref{tab02}.
\begin{table}[h!]
\caption{Average number of distinct groups of particles after time $t=50,100,150,200$.}
\label{tab02}
\begin{tabular}{|c|c|c|c|c|}
\hline
$\phi$ & $n_{t=50}$ & $n_{t=100}$ & $n_{t=150}$ & $n_{t=200}$ \\
\hline\hline
5\%  & 8.3 & 18.1 & 24.0 & 30.0 \\
10\% & 5.7 & 13.3 & 19.2 & 22.0 \\
20\% & 2.9 & 9.0 & 11.3 & 13.3\\
30\% & 2.0 & 5.8 & 8.4 & 10.1\\
40\% & 1.3 & 4.2 & 5.4 & 6.7\\
50\% & 1.0 & 2.1 & 4.4 & 6.2\\
\hline
\end{tabular}
\end{table}

%\newpage
\begin{figure}[h!]
\includegraphics[width=8.5cm]{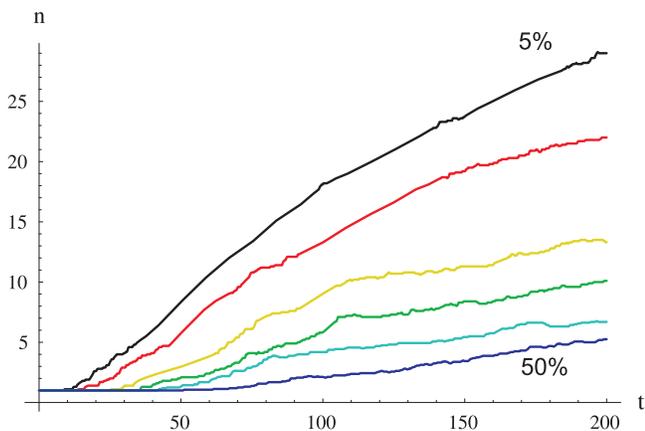}
\caption{Average number $n$ of groups of particles as a function of time. Top-down: $\phi=5\%, 10\%, 20\%,30\%,40\%$ and $50\%$. The errors, calculated as the standard deviation of the mean, are smaller than the distance between the curves.}\label{fig_groups}
\end{figure}

Fig. \ref{fig_groups}
shows the average number $n$ of groups as a function of time, for all studied volume fractions. More dense drops split into a smaller number of groups. Taking into account the mean standard deviation as the error of these results, the dependence on the volume fraction is well established. Once again the difference between the case of $50\%$ and $5\%$ is substantial and suppresses $500\%$. 

The grouping phenomenon is therefore strictly correlated with the initial high concentration of particles. For large volume fractions, hydrodynamic interactions between particles are strong and tend to cluster them. This is why more dense suspension drops destabilize slower. For a high volume fraction, particles stay in one group for a very long time, e.g. for times up to  $\tau=90$ if initially $\phi=50\%$ (refer to Table~\ref{tab02}).

A higher volume fraction leads to a larger suspension viscosity. Therefore, the considered cloud of particles can be interpreted as a drop of a larger viscosity than a host fluid. In the absence of  surface tension, the increase of the drop viscosity leads to its slower and smaller deformation. This result gives an additional information to the numerical study of drop deformations at a finite capillary number, presented in Ref.~\cite{Griggs2007}.

\section{Conclusions}
This paper was devoted to a numerical study of suspension drop evolution in a Poiseuille flow of Stokesian fluid in a parallel-wall channel. The fluid inside the drop was the same as outside. The drop was initially centered on the axes of the channel, away from the walls. We studied the effect of the hydrodynamic interactions between suspended particles on the process of the drop deformation.
Simulations where performed for a wide range of suspension volume fractions, and compared to the evolution of a spherical volume of pure fluid, of the identical initial size and position in the Poiseuille flow as the suspension drop. 

The differences in evolution characteristics between the different volume fractions of suspension drops studied in this paper, are extensive. 
For a low volume fraction, the particles get dispersed evenly occupying an area covering approximately the evolved shape of a comparable volume of pure fluid. 
Dense drops behave differently - for a long time they stay almost non-deformed; later, their evolution is dominated by formation of small groups of rotating particles, which are left behind the drop. 
The clustering is caused by hydrodynamic interactions. Relative motion of close particles tends to hold them together, pushing the flow out of the cluster and hindering its destabilization. The closer the particles are, the longer they stay together in a cluster. For example, at $\phi=50\%$, the stretching  of the initially compact drop, measured as the the dispersion of the particles along the flow, is two times smaller than at $\phi=5\%$, and all the particles stay very close to each other in a compact group more than $6$ times longer. 

The results discussed in this paper have a clear impact on practical application. By changing volume fraction of suspension drops produced in microfluidic devices, or inhalators designed for drug delivery, one is able to control the time-dependent dispersion of the particles entrained by the fluid flow. A large volume fraction will suppress the drop deformation and lead to clustering, while a small one to effective and even spreading of the particles. 

\acknowledgments
This work was supported in part by by the Polish Ministry of Science and Higher Education grant 45/N-COST/2007/0 and the COST P21 Action ``Physics of droplets''.

\appendix
\section{Evolution of an initially spherical fluid-drop}\label{AA}
Volume of a pure-fluid drop remains the same during its evolution. 
The time-dependent circumferential surface of the fluid-drop has been specified in Eq.~\eqref{sd}. 
Therefore, the boundary of the fluid-drop projection on the $y=0$ plane is given by Eq.~\eqref{sd} with $y=0$,
\beq
   (x-\tilde{\text{v}}_0(z) t)^2 = \frac{D^2}{4h^2}-z^2,
   \label{eq:curve_xz}
\eeq
where $\tilde{\text{v}}_0(z)\!=\!4z(1\!-\!z)$. An example of such a shape is shown in Fig.~\ref{fig_evo_fi}.

The circumference of the fluid-drop projection onto the plane $z=1/2$ consists of the following two curves,
\begin{eqnarray}
x_R(y)&=&
   t+\sqrt{\frac{D^2}{4h^2}-y^2}, 
  \label{curve_xy_1}
\end{eqnarray}
i.e. the right boundary of the drop projection, which is simply a half-circle, and 
\begin{eqnarray}
x_L(y)\!\!&=&\!\!\!\!
\min_{z\in \left[\frac{1}{2},\frac{1}{2}+\!\sqrt{\!\frac{D^2}{4h^2}-\!y^2}\right]} \!\left( \! \!\tilde{\text{v}}_0(z) t\!
%\nonumber \\&&
-\!\sqrt{\!\frac{D^2}{4h^2}-\!y^2\!\!-\!\left(\!\!z\!-\frac{1}{2}\!\right)^{\!\!2}}\right)
\label{curve_xz}\nonumber \\
\end{eqnarray}
i.e. the left boundary of the drop projection, which
consists of two parts, 
\bee
    x_{L}(y)\!&\! = \!&\!\!\left\{\!\begin{array}{l}
\!t\!-\sqrt{\frac{D^2}{4h^2}-y^2}\hspace{1.3cm} \mbox{ for } y^2 \!>  \frac{D^2}{4h^2}-\frac{1}{64 t^2},\;\;\;\\\label{L1}\\
\!t\!-\frac{1}{16 t}+\!4 t  \left(\!y^2\!\!-\frac{D^2}{4h^2}\right)\;\,\mbox{ for } y^2 \!\le \frac{D^2}{4h^2}-\frac{1}{64 t^2}.\;\;\;
\end{array}
\right.\nonumber\\\label{curve_xy_22}
\eee
The curve trailing the area occupied by a projected, evolved shape of the spherical volume of fluid is therefore in piece a circle given by Eq. \eqref{curve_xy_1} and the first part of Eq. \eqref{curve_xy_22}, 
 and piecewise a parabola given by the second part of Eq. (\ref{curve_xy_22}). This curve and its derivative are continuous. An example of such a curve is shown in Fig.~\ref{fig_evo_fi_xy}.

%\newpage

\section{Group detection}\label{AB}
\begin{figure}[h!]
\includegraphics[width=8.5cm]{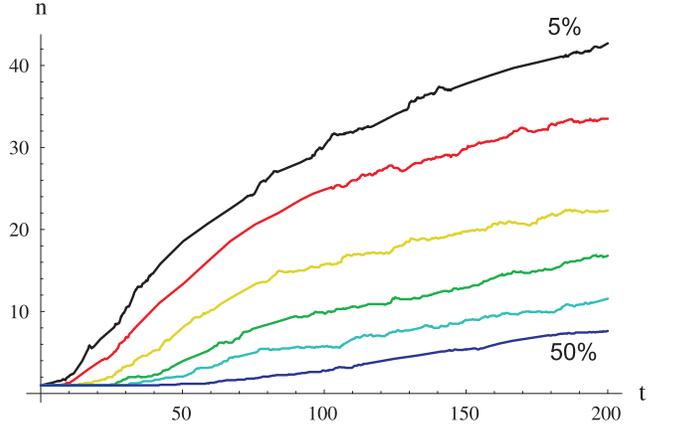}
\caption{Average number $n$ of groups of particles as a function of time. Top-down: $\phi=5\%, 10\%, 20\%, 30\%, 40\%$ and $50\%$. The grouping parameter $g=1.5d$. 
}\label{fig_groups_par}
\end{figure}

The group detection algorithm, as described in section \ref{sec:group} depends on an arbitrary dimensional parameter $g$, which determines the inter-particle distance at which particles are said to belong to separate groups. We chose this parameter to be equal $3d$. Correct conclusions can be drawn from the results of the algorithm, only if a change of $g$ leads only to a quantitative not qualitative change of these results. The number of groups evaluated using the value of $g=1.5d$ (i.e. half of the original value used in Fig.~\ref{fig_groups}) is shown in Fig. \ref{fig_groups_par}. 

%\newpage

\begin{figure}[h!]
\includegraphics[width=8.5cm]{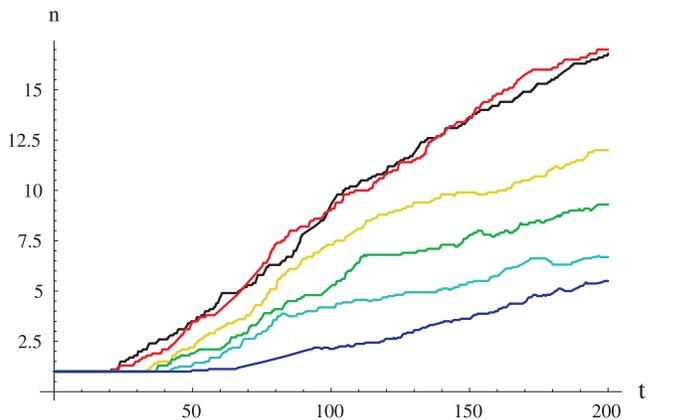}
\caption{Averaged number $n$ of groups of particles as a function of time. $\phi=5\%, 10\%, 50\%$ as indicated, and top-down: $\phi=20\%, 30\%$ and $40\%$. The grouping parameter $g=3h/50$.
}\label{fig_groups_h}
\end{figure}

%\newpage
In Fig. \ref{fig_groups_h}, we have used $g$ scaled with the width of the channel. In this case $g$ has been chosen such that it is $3d$ for $\phi=40\%$ and for all other volume fractions it is scaled proportionally to the (dimensional) distance between the walls $h$,
\beq
 g = \frac{h}{h_{40\%}} 3d = \frac{3}{50} h.
\eeq

In both cases, we find that clearly drops of larger volume fraction tend to form fewer 
clusters. Of course, the number of clusters depends on the choice of the parameter $g$. Nevertheless, the general overall picture of the evolution stays qualitatively the same.


\newpage

\begin{thebibliography}{10}

\bibitem{Sharp2005}
K.~V. Sharp and R.~J. Adrian,
\newblock {Microfluid Nanofluid} {\bf 1}, 376 (2005).

\bibitem{Davis2005}
M.~E. Staben and R.~H. Davis,
\newblock {Int. J. of Multiphase Flow} {\bf 31}, 529 (2005).

\bibitem{Dobson2006}
J.~Dobson,
\newblock {Drug Dev. Res.} {\bf 67} 55 (2006).

\bibitem{Broday2003}
D.~M. Broday and R.~Robinson,
\newblock {Aerosol Science and Technology} {\bf 37}, 510 (2003).

\bibitem{Edwards1997}
D.~A.~Edwards at~al.,
\newblock {Science} {\bf 276}, 1868 (1997).

\bibitem{NitscheBatchelor}
J.~M. Nitsche and G.~K. Batchelor,
\newblock {J. Fluid Mech.} {\bf 340}, 161 (1997).

\bibitem{MMNS} 
G. Machu, W. Meile, L.C. Nitsche and U. Schaflinger,  
\newblock {J. Fluid Mech. } {\bf 447}, 299 (2001).

\bibitem{EMG} M. L. Ekiel-Je\.zewska, B. Metzger, and E. Guazzelli, 
\newblock {Phys. Fluids} {\bf 18}, 038104 (2006).

\bibitem{Janosi1997}
I.~M. Janosi, T.~Tel, D.~E. Wolf, and J.~A.~C. Gallas,
\newblock {Phys. Rev. E} {\bf 56}, 2858 (1997).

\bibitem{Bat07}
G.~K. Batchelor,
\newblock {J. Fluid Mech.} {\bf 56}, 375 (1972).

\bibitem{Jones2001}
R.~B. Jones,
\newblock {J. Chem. Phys.} {\bf 115}, 5319 (2001).

\bibitem{Happel-Brenner:1986}
J.~Happel and H.~Brenner,
\newblock {\em Low {Reynolds} Number Hydrodynamics}.
\newblock Martinus Nijhoff, Dordrecht, 1986.

\bibitem{Kim-Karrila:1991}
S.~Kim and S.~J. Karrila,
\newblock {\em Microhydrodynamics: Principles and Selected Applications}.
\newblock Butterworth-Heinemann, London, 1991.

\bibitem{Bhattacharya-Blawzdziewicz-Wajnryb:2006a}
S.~Bhattacharya, J.~B{\l}awzdziewicz, and E.~Wajnryb,
\newblock {Phys. Fluids} {\bf 18}, 053301 (2006).

\bibitem{Cichocki-Felderhof-Hinsen-Wajnryb-Blawzdziewicz:1994}
B.~Cichocki, B.~U. Felderhof, K.~Hinsen, E.~Wajnryb, and J.~B{\l}awzdziewicz,
\newblock {J. Chem. Phys.} {\bf 100}, 3780 (1994).

\bibitem{Ekiel_Jezewska-Wajnryb:2008}
M.~L. Ekiel-Je{\.z}ewska and E.~Wajnryb,
\newblock {Precise multipole method for calculating hydrodynamic interactions
  between spherical particles in the Stokes flow}.
\newblock In F.~Feuillebois and A~Sellier, editors, {\em {Theoretical Methods
  for Micro Scale Viscous Flows}}, pages 127--172. Transworld Research Network,
  2009.

\bibitem{Bhattacharya-Blawzdziewicz-Wajnryb:2005a}
S.~Bhattacharya, J.~B{\l}awzdziewicz, and E.~Wajnryb,
\newblock {Physica A} {\bf 356}, 294 (2005).

\bibitem{movie03a}
See Supplementary Material Document No.  to watch full simulation of drop evolution for $\phi=5\%$ in MPEG format.

\bibitem{movie03b}
See Supplementary Material Document No.  to watch full simulation of drop evolution for $\phi=10\%$ in MPEG format.

\bibitem{movie03c}
See Supplementary Material Document No.  to watch full simulation of drop evolution for $\phi=20\%$ in MPEG format.

\bibitem{movie03d}
See Supplementary Material Document No.  to watch full simulation of drop evolution for $\phi=30\%$ in MPEG format.

\bibitem{movie03e}
See Supplementary Material Document No.  to watch full simulation of drop evolution for $\phi=40\%$ in MPEG format.

\bibitem{movie03f}
See Supplementary Material Document No.  to watch full simulation of drop evolution for $\phi=50\%$ in MPEG format.

\bibitem{movie04a}
See Supplementary Material Document No.  to watch full simulation of drop evolution for $\phi=5\%$ in MPEG format.

\bibitem{movie04b}
See Supplementary Material Document No.  to watch full simulation of drop evolution for $\phi=10\%$ in MPEG format.

\bibitem{movie04c}
See Supplementary Material Document No.  to watch full simulation of drop evolution for $\phi=20\%$ in MPEG format.

\bibitem{movie04d}
See Supplementary Material Document No.  to watch full simulation of drop evolution for $\phi=30\%$ in MPEG format.

\bibitem{movie04e}
See Supplementary Material Document No.  to watch full simulation of drop evolution for $\phi=40\%$ in MPEG format.

\bibitem{movie04f}
See Supplementary Material Document No.  to watch full simulation of drop evolution for $\phi=50\%$ in MPEG format.



\bibitem{Griggs2007}
A.~J. Griggs, A.~Z. Zinchenko, and R.~H. Davis,
\newblock {Int. J. of Multiphase Flow} {\bf 33}, 182 (2007).

\end{thebibliography}
\end{document}